\documentclass[aps,prl,twocolumn,groupedaddress]{revtex4}

\usepackage{epsfig}
\usepackage{dcolumn}
\usepackage{bm}

\begin{document}

\def\d{{\rm d}}
\def\eps{\varepsilon}
\def\mg{m_{\tilde{g}}}
\def\mba{m_{\tilde{b}_1}}
\def\mbb{m_{\tilde{b}_2}}
\newcommand{\fmslash}[1]{\displaystyle{\not}#1}
\def\lp{\left. }
\def\rp{\right. }
\def\lr{\left( }
\def\rr{\right) }
\def\le{\left[ }
\def\re{\right] }
\def\lg{\left\{ }
\def\rg{\right\} }
\def\lb{\left| }
\def\rb{\right| }
\def\beq{\begin{equation}}
\def\eeq{\end{equation}}
\def\bea{\begin{eqnarray}}
\def\eea{\end{eqnarray}}

\preprint{LPSC 07-096}
\title{SUSY-QCD Corrections to Dark Matter Annihilation in the Higgs Funnel}
\author{Bj\"orn Herrmann}
\author{Michael Klasen}
\email[]{klasen@lpsc.in2p3.fr}
\affiliation{Laboratoire de Physique Subatomique et de Cosmologie,
 Universit\'e Joseph Fourier/CNRS-IN2P3/INPG,
 53 Avenue des Martyrs, F-38026 Grenoble, France}
\date{\today}
\begin{abstract}
We compute the full ${\cal O}(\alpha_s)$ SUSY-QCD corrections to dark matter
annihilation in the Higgs-funnel, resumming potentially large $\mu\tan\beta$
and $A_b$ contributions and keeping all finite ${\cal O}(m_b,s,1/\tan^2
\beta)$ terms. We demonstrate numerically that these corrections strongly
influence the extraction of SUSY mass parameters from cosmological data and
must therefore be included in common analysis tools such as {\tt DarkSUSY}
or {\tt micrOMEGAs}.
\end{abstract}
\maketitle


\section{Introduction}
\label{sec:1}

\vspace*{-72mm}
\noindent LPSC 07-096\\
\vspace*{65mm}

Thanks to the recent mission of the WMAP satellite and other cosmological
observations, the matter and energy decomposition of our Universe is known
today with unprecedented precision \cite{Spergel:2006hy}. Direct evidence
for the existence of Cold Dark Matter (CDM) is accumulating
\cite{Massey:2007wb,Lee:2007}, and its relic density $\Omega_{\rm CDM}$ can
now be constrained to the rather narrow range \cite{Hamann:2006pf}
\beq
 0.094 \,<\, \Omega_{\rm CDM}\,h^2 \,<\, 0.136
 \label{eq:1}
\eeq
at 95\% ($2\sigma$) confidence level. Here, $h$ denotes the present Hubble
expansion rate $H_0$ in units of 100 km s$^{-1}$ Mpc$^{-1}$.

Although the nature of Cold Dark Matter still remains unknown, it is likely
to be composed of Weakly Interacting Massive Particles (WIMPs), as proposed
by various extensions of the Standard Model (SM) of particle physics. In
Supersymmetry (SUSY), a natural candidate is the Lightest Supersymmetric
Particle (LSP), which is stable, if $R$-parity is conserved. It is usually
the lightest of the four neutralinos, denoted $\tilde{\chi}_1^0$ or shortly
$\chi$.

The Minimal Supersymmetric Standard Model (MSSM) depends a priori on 124
soft SUSY-breaking parameters, which are often restricted to five universal
parameters that are imposed at the unification scale and can be constrained
using data from high-energy colliders. As the lightest neutralino relic
density depends also on these parameters, its computation is another
powerful tool to put constraints on the parameter space and provide
complementary information, in particular at high energies or masses that
would otherwise not be accessible at colliders.

To evaluate the number density $n$ of the relic particle with velocity $v$,
one has to solve the Boltzmann equation 
\beq
 \frac{dn}{dt} = -3 H n - \langle\sigma_{\rm eff}\,v \rangle
 \left( n^2 - n_{eq}^2 \right)
 \label{eq:2}
\eeq
with the Hubble rate $H$ and the thermal equilibrium density $n_{eq}$. The
present number density $n_0$ is directly related to the relic density
$\Omega_{\rm CDM}h^2=m_{\chi}n_0/\rho_c\propto\langle\sigma_{\rm eff}\,v
\rangle^{-1}$, where $m_\chi$ is the LSP mass, $\rho_{c}=3H_0^2/(8\pi G_N)$
is the critical density of our Universe, and $G_N$ is the gravitational
constant \cite{Bertone:2004pz}. The effective cross section $\sigma_{\rm
eff}$ involves all annihilation and co-annihilation processes of the relic
particle $\chi$ into SM particles, and $\langle\sigma_{\rm eff}v \rangle$ in
Eq.\ (\ref{eq:2}) signifies the thermal average of its non-relativistic
expansion ($v\ll c$).

The most important processes contributing to $\sigma_{\rm eff}$ are those
that have two-particle final states and that occur at the tree level
\cite{Jungman:1995df}. Possible final states include those with
fermion-antifermion pairs as well as with combinations of gauge ($W^{\pm}$,
$Z^0$) and Higgs ($h^0$, $H^0$, $A^0$, $H^{\pm}$) bosons, depending on the
region of the parameter space. Processes producing fermions or antifermions
may be detectable either directly or through their annihilation into
photons. In addition, these channels are always open (for $b$-quarks if
$m_\chi\geq4.5$ GeV) in contrast to the other channels, which may
be suppressed or even closed \cite{Jungman:1995df,Bertone:2004pz}.

Several public codes perform a calculation of the dark matter relic density
within supersymmetric models. The most developped and most popular ones are
{\tt DarkSUSY} \cite{Gondolo:2004sc} and {\tt micrOMEGAs}
\cite{Belanger:2001fz}. All relevant processes are implemented in these
codes, but for most of them no (or at least not the full) higher order
corrections are included. However, due to the large magnitude of the strong
coupling constant, QCD and SUSY-QCD corrections are bound to affect the
annihilation cross section in a significant way. They may even be enhanced
logarithmically by kinematics or in certain regions of the parameter space.

In this Letter, we compute these corrections for neutralino-pair
annihilation
into a bottom quark-antiquark pair through the $s$-channel exchange of a
pseudoscalar Higgs-boson $A^0$. This process dominates in the so-called
{\it A-funnel} region of minimal supergravity (mSUGRA) parameter space at
large $\tan\beta$, which is theoretically favored by the unification of
Yukawa couplings in Grand Unified Theories (GUTs) \cite{Carena:1994bv}.
Supposing a WIMP mass of 50-70 GeV, this process has also been claimed to
be compatible with the gamma-ray excess observed in all sky directions by
the EGRET satellite \cite{deBoer:2005tm}. However, the corresponding
scenarios may lead to antiproton overproduction, so that they would not be
compatible with the observed antiproton flux \cite{Bergstrom:2006tk}.

\section{Analytical results}
\label{sec:2}

Denoting the neutralino and $b$-quark velocities by 
\bea
 \beta_{\chi} = {v\over2} = \sqrt{1-\frac{4m_{\chi}^2}{s}} &{\rm and}&
 \beta_b = \sqrt{1-\frac{4m_b^2}{s}}
 \label{eq:3}
\eea
and the squared total center-of-mass energy by $s$,
the properly antisymmetrized neutralino annilation cross section
can be written at leading order (LO) of perturbation theory as
\bea
 \sigma_{\rm LO}\,v &=& {1\over2}{\beta_b\over8\pi s} 
 \frac{N_C\,g^2 \,T_{A11}^2 \,h_{Abb}^2 \,s^2}{\left|
 s-m_A^2+im_A \Gamma_A \right|^2}.
 \label{eq:4}
\eea
It is proportional to the inverse of the flux factor $sv$, the integrated
two-particle phase space $s\beta_b/(8\pi s)$, the number of quark colors
$N_C=3$ and the squares of the weak coupling constant $g$, a neutralino
mixing factor
\bea
 T_{Aij}&=& \frac{1}{2} \Big(N_{2j}-\tan\theta_W N_{1j}\Big) 
                        \Big(N_{4i}\cos\beta-N_{3i}\sin\beta\Big)\nonumber\\
 &&+ (i\leftrightarrow j),
 \label{eq:5}
\eea
the bottom-quark mass $m_b$ through the Yukawa coupling $h_{Abb}=-gm_b\tan
\beta/(2 m_W)$ and the Higgs-boson propagator. Expanding in powers of $v^2$,
we obtain $\sigma_{\rm LO} \,v \doteq a_{\rm LO} + b_{\rm LO} \,v^2 +
{\cal O}\left( v^4 \right)$ with
\bea
 \!\!a_{\rm LO} \!\!&=&\!\! 2b_{\rm LO}\! =
 \!\frac{N_C\,g^2\, T_{A11}^2\, h_{Abb}^2}
 {4\pi m_{\chi}^2 \left| 4-\frac{m_A^2}{m_{\chi}^2}
 +{im_A \Gamma_A\over m_\chi^2}\right|^2} \sqrt{1-\frac{m_b^2}{m_{\chi}^2}}
 ~~
 \label{eq:6}
\eea
in agreement with Ref.\ \cite{Jungman:1995df}.

\begin{figure}
 \centering
 \epsfig{file=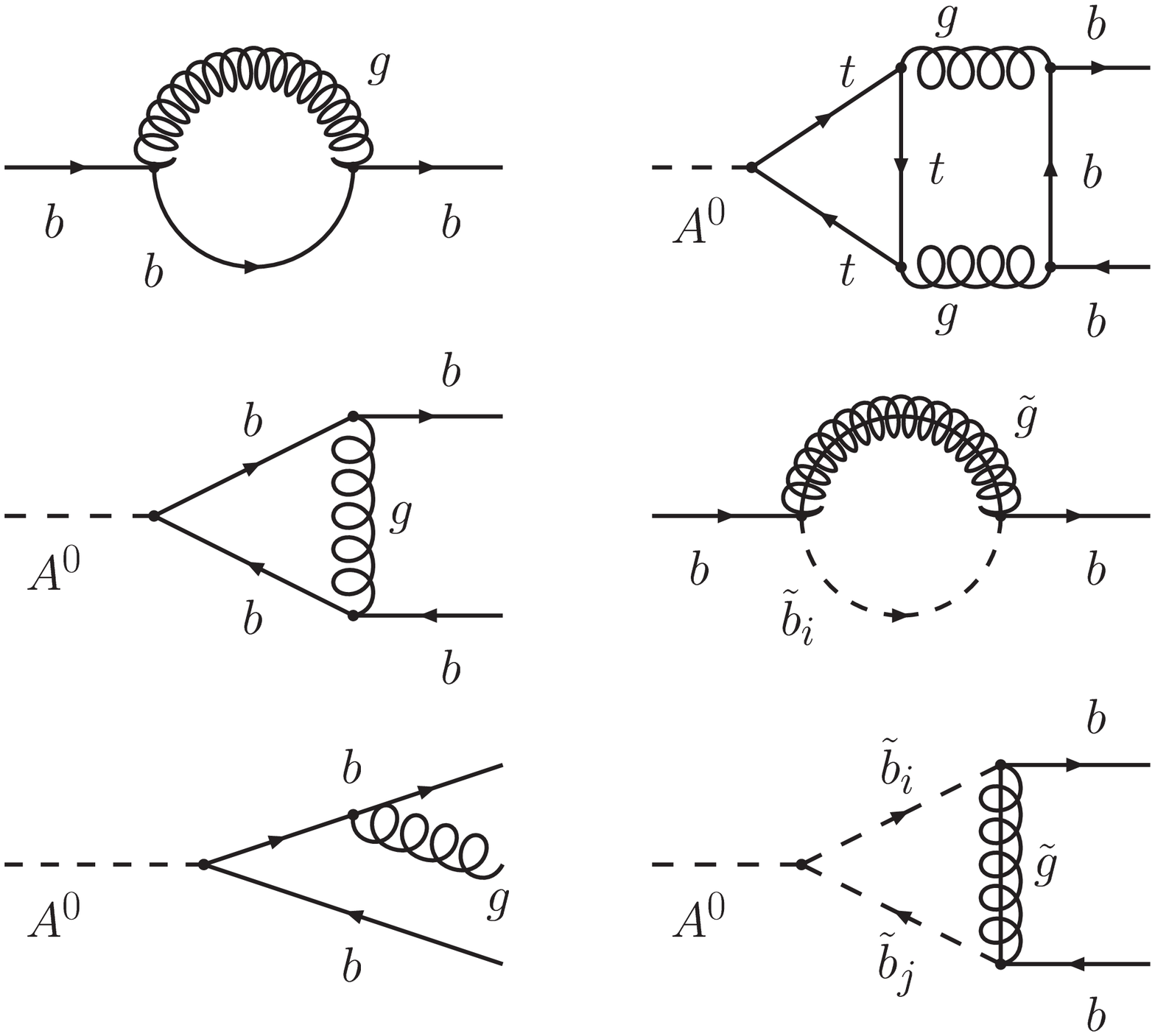,width=.76\columnwidth}
 \caption{\label{fig:1}Left: Diagrams for self-energy (top), vertex
 (center), and real (bottom) QCD corrections. Right: Diagrams for two-loop
 top-quark (top) and one-loop self-energy (center) and vertex (bottom)
 SUSY-QCD corrections to the process $\chi\chi\to A^0\to b\bar{b}$.}
\end{figure}

Using standard methods for the virtual one-loop and real emission
contributions shown on the left-hand side of Fig.\ \ref{fig:1}, we compute
%
%
the ${\cal O}(\alpha_s)$ QCD correction in the {\em on-shell scheme}
\bea
 \Delta_{\rm QCD}^{(1)} \!\!&=&\!\! \lr\frac{\alpha_s(s)}{\pi}\rr C_F
 \Biggr[ \frac{1+\beta_b^2}{\beta_b}\Biggr( 4 \text{Li}_2\frac{1-\beta_b}
 {1+\beta_b}
 + 2 \text{Li}_2\frac{\beta_b-1}{1+\beta_b} \nonumber\\
 &&- 3 \log\frac{2}{1+\beta_b}\log\frac{1+\beta_b}{1-\beta_b}
   - 2 \log\beta_b \log\frac{1+\beta_b}{1-\beta_b} \Biggr)\nonumber\\
 &&- 3\log\frac{4}{1-\beta_b^2}-4\log\beta_b
   + \frac{3}{8}\left( 7-\beta_b^2 \right) \nonumber\\
 &&+ \frac{1}{16\beta_b}\left( 19+2\beta_b^2+3\beta_b^4\right)\log
     \frac{1+\beta_b}{1-\beta_b}\Biggr],
\label{eq:7}
\end{eqnarray}
which agrees with the known result for pseudoscalar Higgs-boson decays
\cite{Drees:1989du} and contributes to the total correction
\bea
 \sigma &=& \sigma_{\rm LO} \Big[ 1 + \Delta_{\rm QCD} + \Delta_{\rm top}
 + \Delta_{\rm SUSY} \Big]
\eea
and equivalently for $\Gamma_A$. Here, $\Delta_{\rm QCD}=
({\alpha_s\over\pi})  \Delta_{\rm QCD}^{(1)}+
({\alpha_s\over\pi})^2\Delta_{\rm QCD}^{(2)}+
({\alpha_s\over\pi})^3\Delta_{\rm QCD}^{(3)}+...$

In the limit $m_b^2\ll s$ ($\beta_b\to1$), the correction in Eq.\
(\ref{eq:7}) develops a logarithmic mass singularity
\bea
 \Delta_{\rm QCD}^{(1)} &\simeq& \lr\frac{\alpha_s(s)}{\pi}\rr C_F 
 \le-{3\over2}\log{s\over m_b^2}+{9\over4}\re,
\eea
which can be resummed to all orders using the renormalization group, i.e.\
by replacing $m_b$ with the running mass $\bar{m}_b(s)$ in the Yukawa
coupling $h_{Abb}$ \cite{Braaten:1980yq}. The remaining finite
QCD-corrections {\it in the $\overline{\it MS}$-scheme} are known up to
${\cal O}(\alpha_s^3)$ \cite{Chetyrkin:1996sr}
\bea
 \Delta_{\rm QCD} \!\!&=&\!\!
 \lr\frac{\alpha_s(s)}{\pi}\rr   C_F {17\over4}
 \!+\!\lr\frac{\alpha_s(s)}{\pi}\rr^2 \!\!(35.94-1.36 n_f)\nonumber\\
 &+&\!\!\lr\frac{\alpha_s(s)}{\pi}\rr^3 (164.14-25.76 n_f+0.259 n_f^2).
\eea

A separately gauge-independent ${\cal O}(\alpha_s^2)$ correction is induced
by the top-quark loop diagram shown on the upper right-hand side of Fig.\
\ref{fig:1}. Its contribution \cite{Chetyrkin:1995pd}
\beq
 \Delta_{\rm top} = \frac{1}{\tan^2\beta} \lr\frac{\alpha_s(s)}{\pi}\rr^2
 \left[ \frac{23}{6}-\log\frac{s} {m_{t}^2} + \frac{1}{6}\log^2
 \frac{\bar{m}_{b}^2(s)}{s} \right]
 \label{eq:11}
\eeq
can be important at small values of $\tan\beta$, but is largely suppressed
in the Higgs-funnel region considered here.

In SUSY, additional ${\cal O} (\alpha_s)$ corrections arise through the
sbottom-gluino exchanges shown in the central and lower right-hand side
diagrams of Fig.\ \ref{fig:1}. The self-energy diagram leads to the mass
renormalization \cite{Carena:1994bv}
\beq
 \Delta m_b= \lr\frac{\alpha_s(s)}{\pi}\rr C_F{\mg\over2}(A_b-\mu\tan\beta)
 I(\mba^2,\mbb^2,\mg^2),
\eeq
which is for $m_b\ll m_{\rm SUSY}$ proportional to
\bea
 \sin2\theta_b&=&{2m_b(A_b-\mu\tan\beta)\over\mba^2-\mbb^2},
\eea
i.e.\ the off-diagonal component of the sbottom mass matrix, and the 3-point
function at zero external momentum
\bea
 I(a,b,c)&=&{ab\log{a\over b}+bc\log{b\over c}+ca\log{c\over a}\over
 (a-b)(b-c)(c-a)}.
\eea
In this low-energy (LE) limit, and neglecting $A_b$ with respect to the
$\tan\beta$-enhanced $\mu$, the vertex correction equals the mass
renormalization \cite{Carena:1999py} up to a factor $1/\tan^2\beta$, so that
the total SUSY correction becomes
\bea
 \Delta_{\rm SUSY}^{\rm (LE)}&=& \lr\frac{\alpha_s(s)}{\pi}\rr C_F
 \lr 1+{1\over\tan^2\beta}\rr\mg\mu\tan\beta\nonumber\\
 &\times&I(\mba^2,\mbb^2,\mg^2).
\eea

It has long been known that for large $\tan\beta$, $\Delta m_b$ can be
significant and must be resummed by replacing $m_b\to m_b/(1+\lim_{A_b\to0}
\Delta m_b)$ in the Yukawa coupling $h_{Abb}$ \cite{Carena:1999py}. More
recently it has been observed that $A_b$ may be of similar size as $\mu\tan
\beta$, e.g.\ in no-mixing scenarios, so that its contribution must also be
resummed by replacing $\lim_{A_b\to0}\Delta m_b \to (\lim_{A_b\to0}\Delta
m_b)/(1+\lim_{\mu\tan\beta\to0}\Delta m_b)$ \cite{Guasch:2003cv}. Our result
for the full SUSY-QCD correction $\Delta_{\rm SUSY}$ agrees with those in
\cite{Dabelstein:1995js,Coarasa:1995yg}, and we implement the finite
${\cal O}(m_b,s,1/\tan^2\beta)$ remainder as described in
\cite{Guasch:2003cv}.

\section{Numerical evaluation}
\label{sec:3}

For our numerical study of the impact of QCD, top-quark loop, and SUSY-QCD
corrections on dark matter annihilation in the Higgs-funnel, we place
ourselves in a minimal supergravity (mSUGRA) scenario with $A_0=0$ and large
$\tan\beta=44.5$ $(54)$ for $\mu<0$ ($\mu>0$), which still allows for
electroweak symmetry breaking (EWSB) in a large region of the scanned
$m_{1/2}-m_0$ plane. The weak-scale MSSM parameters are then determined
with {\tt SPheno} \cite{Porod:2003um}, which includes resummed $\lim_{A_b\to
0}\Delta m_b$ corrections, and the physical Higgs and SUSY masses with
{\tt FeynHiggs} \cite{Heinemeyer:1998yj} after imposing the current SM
masses (in particular $\bar{m}_b(m_b)=4.2$ GeV and $m_t=174.2$ GeV), gauge
couplings ($\alpha$ and $\sin^2\theta_W$ in the improved Born approximation,
$\alpha_s(M_Z)=0.1176$), and direct and indirect SUSY mass limits
\cite{Yao:2006px}. For the impact of SUSY spectrum calculations on dark
matter annihilation see \cite{Belanger:2005jk}.

\begin{figure}
 \centering
 \epsfig{file=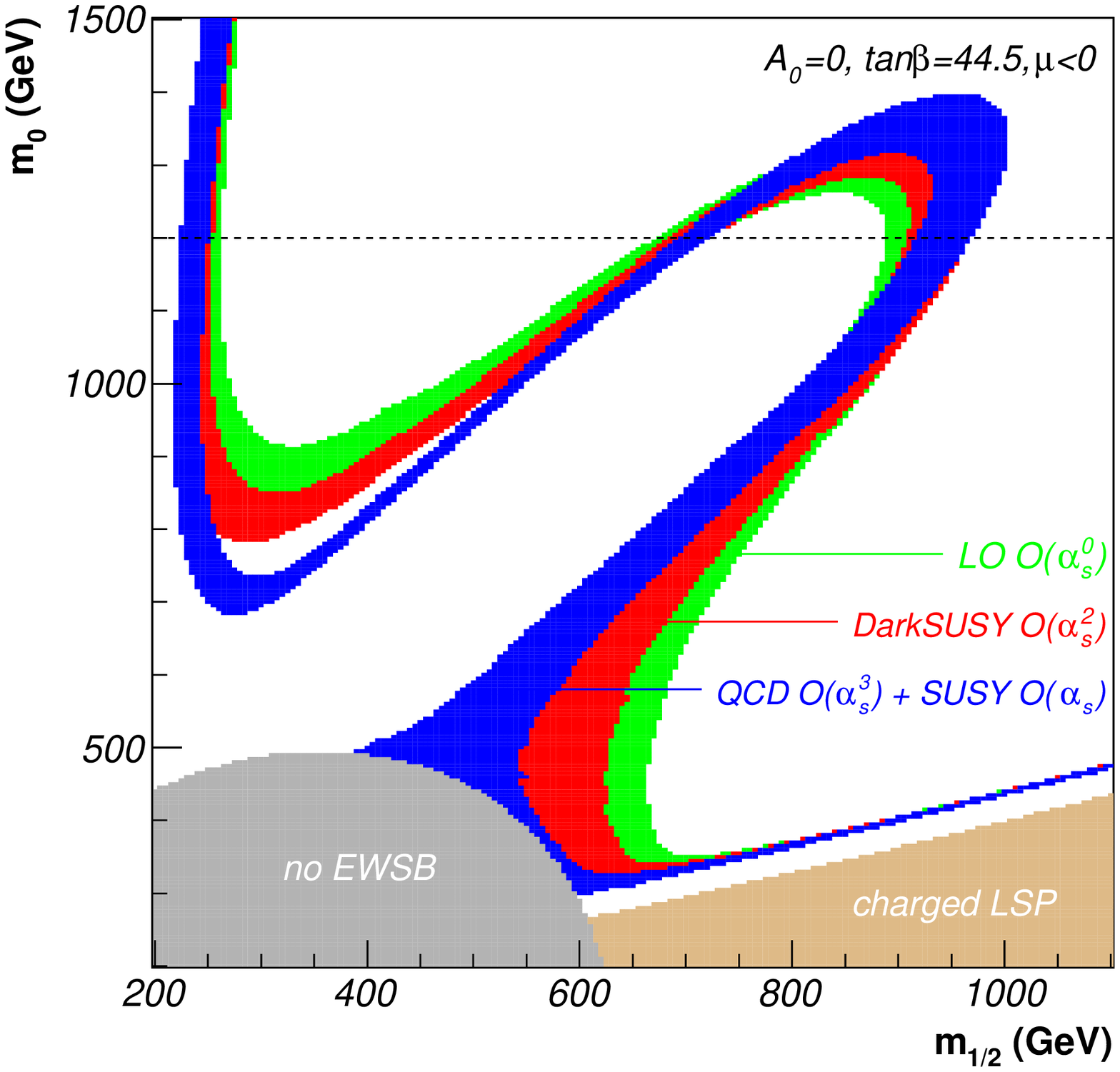,width=0.76\columnwidth}
 \epsfig{file=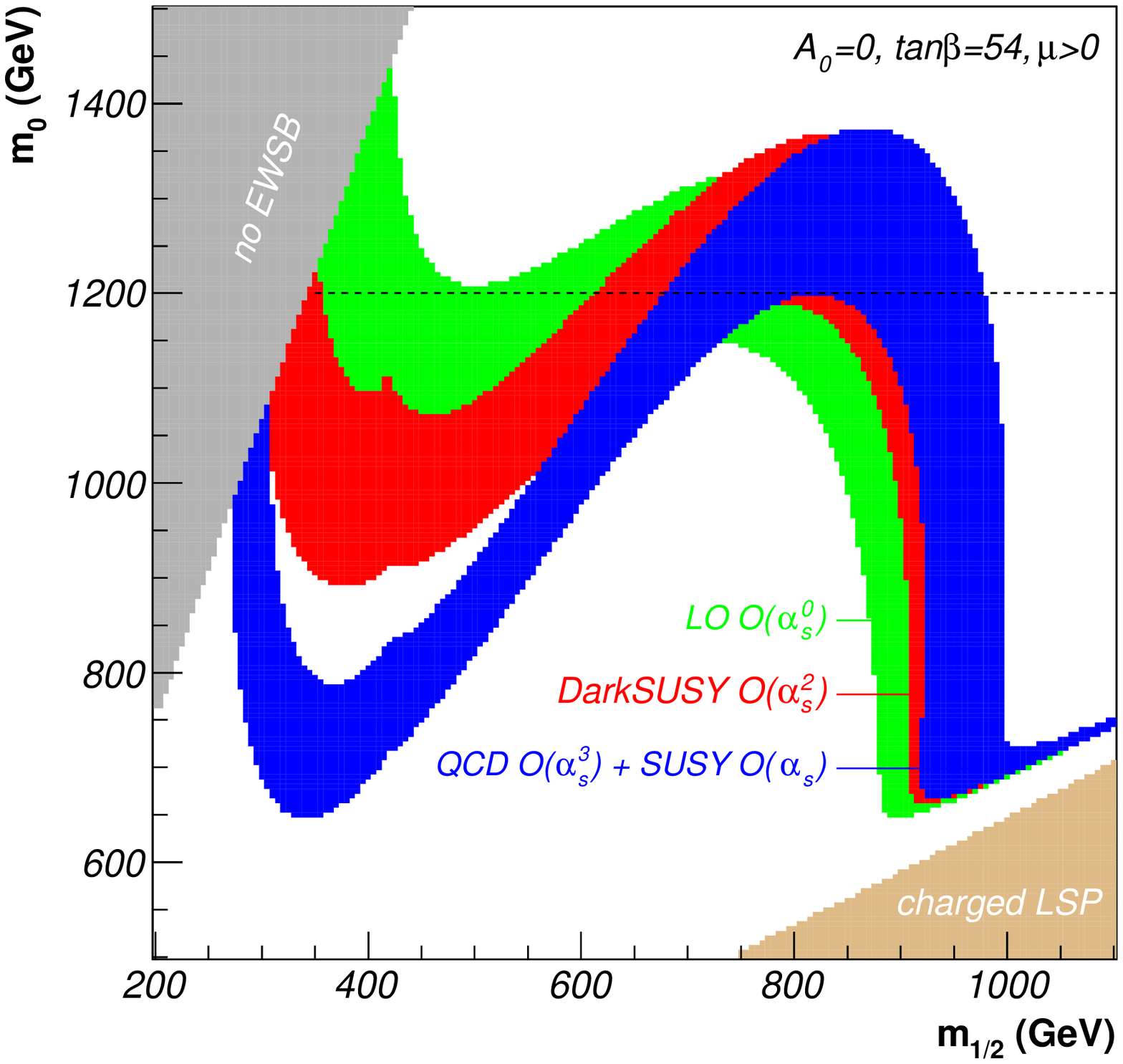,width=0.76\columnwidth}
 \caption{\label{fig:2}Regions in the mSUGRA $m_{1/2}-m_0$ plane forbidden
 by a charged LSP/no EWSB and favored by the observed $\Omega_{\rm CDM}\,
 h^2$ for large $\tan\beta=44.5$ $(54)$ and $\mu<0$ ($\mu>0$).}
\end{figure}

Comparing the observed CDM relic density in Eq.\ (\ref{eq:1}) to the one
calculated with {\tt DarkSUSY} \cite{Gondolo:2004sc}, which includes the QCD
corrections up to ${\cal O}(\alpha_s^2)$ and where we have added the
${\cal O}(\alpha_s^3)$ QCD and ${\cal O}(\alpha_s)$
SUSY-QCD corrections described above, we determine the allowed regions in
the $m_{1/2}-m_0$ plane shown in Fig.\ \ref{fig:2}. The
%
%
Higgs-funnel contribution to $\sigma_{\rm eff}$ rises from 40\% for low
values of $m_{1/2}$ (or $m_0$ for $\mu<0$) to more than 95\%, when $m_{1/2}$
(and $m_0$) is (are) large. For $m_0=1200$ GeV (dashed line), the condition
$(m_A-2m_\chi)/\Gamma_A=0$ is, e.g., satisfied when $m_{1/2}=840$ (970) GeV
for $\mu<0$ ($\mu>0$), where $m_\chi\simeq 360$ (420) GeV. It is obvious
from Fig.\ \ref{fig:2} that the LO allowed regions (light) are dramatically
changed by the ${\cal O}(\alpha_s^2)$ (medium) and ${\cal O}(\alpha_s^3)$
QCD and ${\cal O}(\alpha_s)$ SUSY-QCD corrections (dark), which reduce
$\sigma_{\rm eff}$ by more than a factor of two. The increase in
$\Omega_{\rm CDM}$ must therefore be compensated by smaller masses.
However, $\Gamma_A$ is reduced by appoximately the same amount, so that on
the Higgs pole, where $\Gamma_A$ is of particular importance, the effect is
reversed. As expected, the effect is negligible in the focus point (very low
$m_{1/2}$) and co-annihilation (very low $m_0$, see also
\cite{Freitas:2007sa}) regions.

\begin{figure}
 \centering
 \epsfig{file=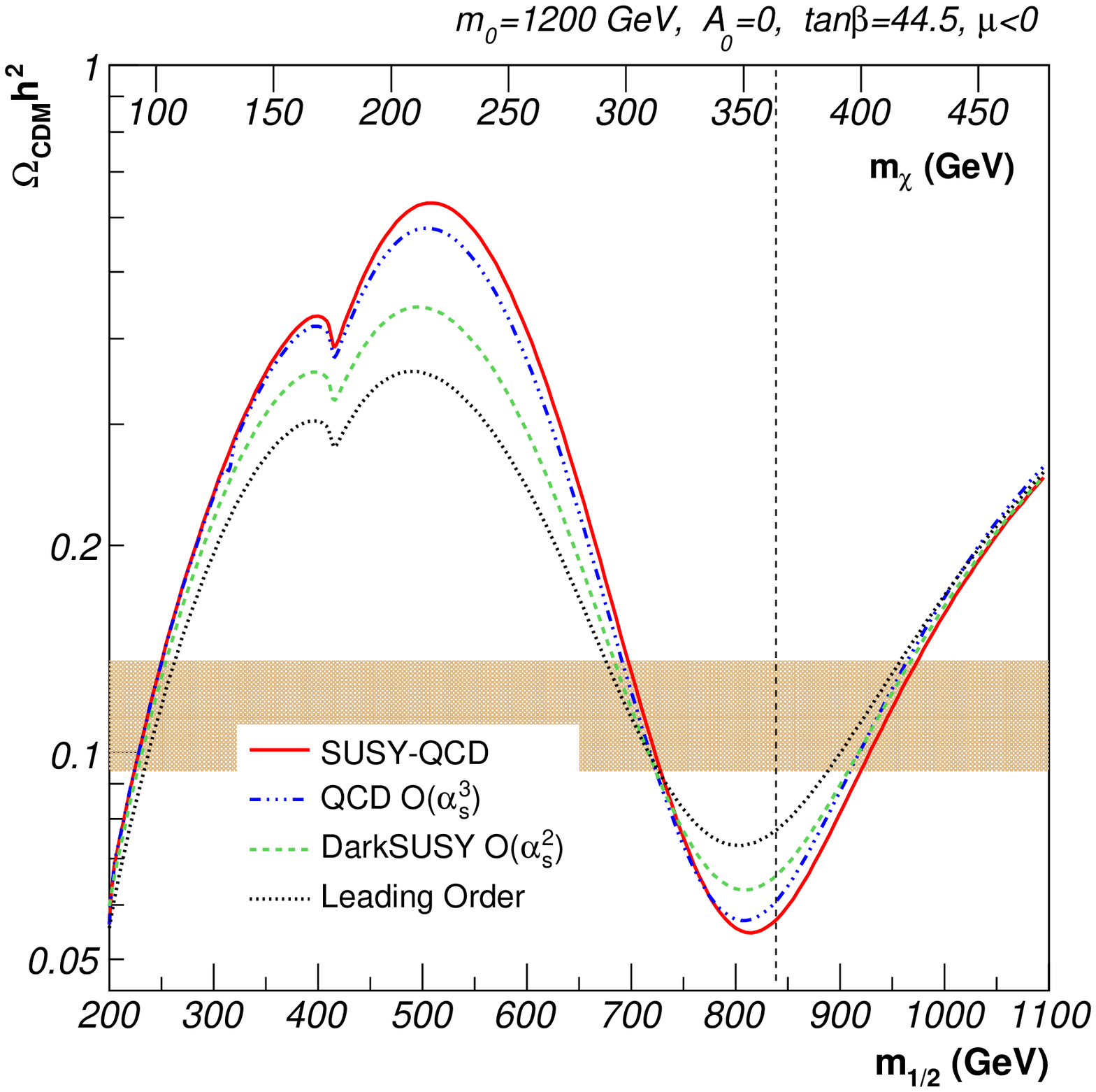,width=0.76\columnwidth}
 \epsfig{file=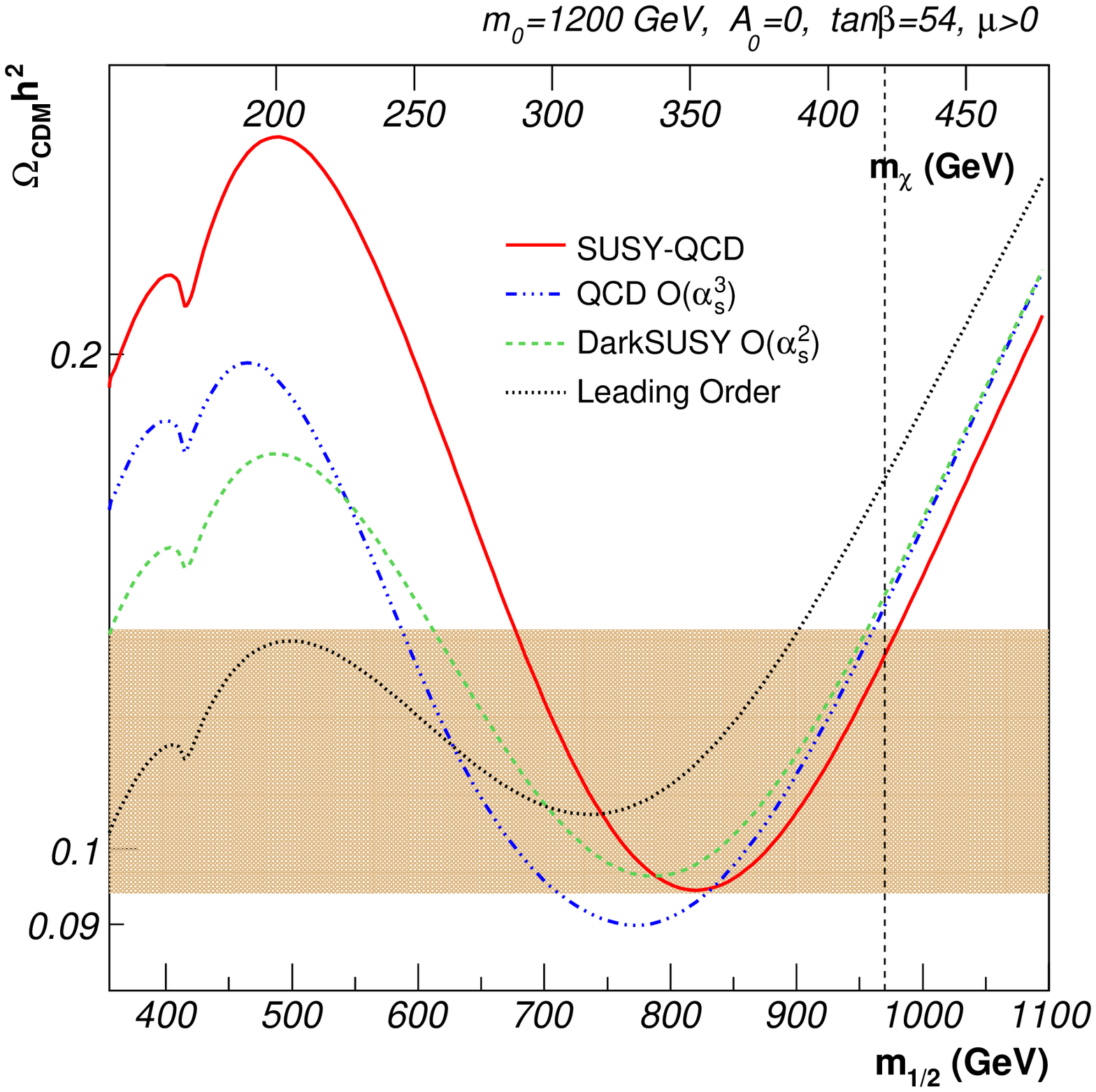,width=0.76\columnwidth}
 \caption{\label{fig:3}LO (dotted), ${\cal O}(\alpha_s^2)$ (dashed) and
 ${\cal O}(\alpha_s^3)$ (dot-dashed) QCD and ${\cal O}(\alpha_s)$ SUSY-QCD
 (full) corrected predictions for $\Omega_{\rm CDM}\,h^2$ compared to the
 experimentally allowed range as a function of $m_{1/2}$. The position of
 the Higgs pole is indicated by a vertical dashed line.}
\end{figure}

In Fig.\ \ref{fig:3} we plot the relic density for $m_0=1200$ GeV as a
%
%
function of $m_{1/2}$. Since $\Gamma_A$ increases with $m_{1/2}$, in
particular for $\mu>0$, $\sigma_{\rm eff}$ ($\Omega_{\rm CDM}$) reaches its
maximum (minimum) at some distance from the pole (vertical dashed line).
Further away, the (SUSY-)QCD corrections decrease (increase)
$\sigma_{\rm eff}$ ($\Omega_{\rm CDM}$) as expected. Closer to the pole,
the reduced width becomes important, so that the maximum (minimum)
approaches the pole. The effect of the ${\cal O}(\alpha_s^2)$
QCD corrections already included in {\tt DarkSUSY} and {\tt micrOMEGAs}
\cite{Gomez:2002tj,Allanach:2004xn} is considerably enhanced by our
${\cal O}(\alpha_s^3)$ QCD and SUSY-QCD corrections, while the top-quark
loop contributes less than 0.01\% for the large values of $\tan\beta$ under
consideration here. Note that the dip at $m_{1/2}=420$ GeV occurs when
$m_\chi=m_t$, where the $t\bar{t}$ annihilation channel is opened
\cite{Moroi:2006fp}.

\section{Conclusion}
\label{sec:4}

In summary, we have computed the full ${\cal O}(\alpha_s)$ SUSY-QCD
corrections to dark matter annihilation in the Higgs-funnel, resumming
potentially large $\mu\tan\beta$ and $A_b$ contributions and keeping
all finite ${\cal O}(m_b,s,1/\tan^2\beta)$ terms. We have demonstrated
numerically that these corrections strongly influence the extraction of
SUSY mass parameters from cosmological data and must therefore be included
in common analysis tools like {\tt DarkSUSY} or {\tt micrOMEGAs}.


\acknowledgments
We thank S.\ Kraml, M.\ Steinhauser and T.\ Tait for valuable discussions.
This work has been supported by a Ph.D.\ fellowship of the French ministry
for education and research.


\end{document}